\documentstyle[12pt,fleqn]{article}

\addtocounter{section}{-1}
\def\beeq{\begin{equation}}
\def\eneq{\end{equation}}
\def\beeqa{\begin{eqnarray}}
\def\eneqa{\end{eqnarray}}

\def\eg{E_{\rm g}}
\def\soc{{\rm C}_{60}}
\def\rug{{\rm C}_{70}}

\begin{document}
{}~
\vspace{1.4cm}
\begin{flushleft}
\underline{DIMERIZATION AND ENERGY-LEVEL STRUCTURES IN FULLERENE TUBULES}\\
\underline{INVESTIGATED WITH AN ELECTRON-PHONON MODEL}\\
\vspace{1.4cm}
Kikuo H{\sc arigaya}\\
Fundamental Physics Section, Electrotechnical
Laboratory, Tsukuba, Ibaraki 305 (Japan)\\
Mitsutaka F{\sc ujita}\\
Institute of Materials Science,
University of Tsukuba, Tsukuba, Ibaraki 305 (Japan)
\end{flushleft}
\vspace{2.3cm}
\noindent
{ABSTRACT}

Possible dimerization patterns and electronic structures in fullerene
tubules as the $\pi$-con\-ju\-gated systems are studied with the extended
Su-Schrieffer-Heeger model.  We assume various lattice geometries,
including helical and nonhelical tubules, and tubules with end caps.
The model is solved for the half-filling case of $\pi$-electrons.
(1) When the undimerized systems do not have a gap, the \underline{Kekul\'{e}
structures} tend to occur.  These structures are commensurate with
the boundary condition in the direction perpendicular to the tubular
axis.  The energy gap is of the order of the room temperatures at most.
Thus, the \underline{nearly metallic properties} would be expected.
(2) If the undimerized systems have a large gap ($\sim$ 1eV), the
most stable structures are the \underline{chain-like distortions} where the
direction of the arranged {\sl trans}-polyacetylene chains is along
almost the tubular axis.  When we try to obtain the Kekul\'{e}
structures, the mismatch of the boundary condition occurs and they
are energetically unfavorable.  The electronic structures are of
\underline{semiconductors} due to the large gap.

\vspace{0.5cm}
\noindent
{INTRODUCTION}

Recently, a new form of carbons, ``fullerene tubules",
has been synthesized [1].
A tubule has the structure like a cylinder made from a graphite sheet.
The typical diameter of the tubules is of the order of one nanometer.
The maximum length of the tubules is more than one micrometer.
Thus, a tubule can be regarded as a new one-dimensional conductor.

The electronical properties have been theoretically calculated [2].
A tubule can show metallic or semiconductor-like properties,
depending upon its geometry.  In these studies [2], all of the
carbon atoms have been assumed to be equivalent.  The possibility
of the \underline{bond alternation patterns} (namely the dimerizations)
has not been considered yet.   The occurrence of dimerizations would be
quite interesting when we regard tubules as quasi one-dimensional
$\pi$-conjugated systems.

In this report, we extend the
Su-Schrieffer-Heeger (SSH) model [3] of conjugated
polymers in order to apply to fullerene tubules.  The electronic properties and
the possible dimerization patterns are analyzed by using the finite-size
scaling method [4].  We will obtain the Kekul\'{e} structure for the
nearly-metallic tubules, and the chain-like distortion for the
semiconductoring tubules.

\vspace{0.5cm}
\noindent
{MODEL}

The following extended SSH hamiltonian [4] is applied to various geometries
of fullerene tubules:
\beeq
H = \sum_{\langle i,j \rangle, \sigma} ( - t_0 + \alpha y_{i,j} )
( c_{i,\sigma}^\dagger c_{j,\sigma} + {\rm h.c.} )
+ \frac{K}{2} \sum_{\langle i,j \rangle} y_{i,j}^2.
\eneq
Here, $c_{i,\sigma}$ is an annhilation operator of a $\pi$-electron; the
quantity $t_0$ is the hopping integral of the ideal undimerized system;
$\alpha$ is the electron-phonon coupling; $y_{i,j}$ indicates the bond
variable which measures the length change of the bond between the $i$- and
$j$-th sites from that of the undimerized system; the sum is
taken over nearest neighbor pairs $\langle i j \rangle$; the
second term is the elastic energy of the lattice; and the
quantity $K$ is the spring constant.  The model with the given geometry
of the tubule is solved with the assumption of the adiabatic approximation
and by the iteration method.  The periodic boundary condition is called
for also.

The value $t_0 = 2.5$eV is taken from that of graphite and polyacetylene [4].
We use $\alpha = 6.31$eV/\AA and $K=49.7$eV/\AA$^2$ [4]. The dimensionless
electron-phonon coupling $\lambda \equiv 2\pi \alpha^2 / \pi K t_0$,
analogous to that in superconductivity, has the weak coupling value 0.20.

\vspace{0.5cm}
\noindent
{HELICAL AND NONHELICAL TUBULES}

Figure 1 shows the way of making general tubules and the notations.  The
lattice points in the honeycomb lattice are labeled by the vector
$(m,n) \equiv m {\bf a} + n {\bf b}$, where ${\bf a}$ and
${\bf b}$ are the unit vectors.  Any structure of tubules can be
produced by combining the origin $(0,0)$ with one of $(m,n)$.
This vector can be used as a name of each tubule.  When the
electron-phonon coupling does not exist, i.e., $\lambda = 0$,
there are two possibilities of electronic structures: metals
and semiconductors.  When the origin of the honeycomb lattice
pattern is so combined with one of the open circles
as to make a tubule, the metallic properties will be expected
because of the presence of the Fermi surface.  This case corresponds
to the vectors where $m-n$ is a multiple of three.  If the origin
is combined with the filled circles, there remains a large gap
of the order of 1eV.  The system will be a semiconductor.  The
similar properties have been discussed in several recent papers [2].

When there is a non-zero electron-phonon coupling, several kinds of bond
ordered configurations can be expected.  The most characteristic
pattern in the two dimensional graphite plane is the namely Kekul\'{e}
structure.  The pattern is superposed with the honeycomb
lattice in Fig. 1.  The short and long bonds are indicated
by the thick and normal lines, respectively.  This pattern is
commensurate with the lattice structure for the tubules with $m-n$
being multiples of three.  Therefore, the Kekul\'{e} structure
is one of the candidates of stationary solutions.  For other tubules,
the Kekul\'{e} pattern is forbidden due to the structural
origin.  In contrast, patterns, where {\sl trans}-polyacetylene--chains
are connected by the long bonds in the transverse direction, can
be realized for any set of $(m,n)$.  This will be discussed in relation
with Figs. 4 and 5.

We shall explain our strategy of the investigation.  When the origin $(0,0)$
is combined with $(5,5)$, we obtain a nonhelical tubule.  Ten carbons
are arranged in a ring along the tubule axis.
When we make the tubule $(6,4)$ furthermore, we cut all the
rings in the tubule $(5,5)$ and connect the neighboring rings each
other.  The tubule becomes helical. If we connect next nearest
neighboring rings, we obtain the tubule $(7,3)$.  The tubule becomes
more helical.  When we repeat the above procedure further, we
obtain the nonhelical tubule at (10,0).  We pursuit changes in
dimerization patterns and electronic energy levels, starting from
the tubule $(5,5)$.

\vspace{0.5cm}
\noindent
{KEKUL\'{E} STRUCTURE}

We have actually obtained the \underline{Kekul\'{e} structure} as the
most stable solution for the tubules, (5,5) and (8,2).  In these tubules,
the Kekul\'{e} pattern is commensurate with the geometrical structure.
We discuss variations of the energy gap and the bond variables as
functions of the carbon number $N$.

Figure 2 shows the energy gap $\eg$ of the tubule (5,5).
The gap varies linearly as a function of $1/N$ due to the one
dimensional nature.  When $N \sim 100$, $\eg$ is of the order of
1eV.  When $N \sim 500$, it becomes of the order of 0.1eV.
The extrapolated value at $N \rightarrow \infty$ is $4.39
\times 10^{-3}$eV.  This is apparently lower than the room temperature.
In addition, we should pay attention to the thermal fluctuation of phonons.
Thus, we can expect \underline{nearly metallic behaviors} even
at low temperatures.

Figure 3 displays the $1/N$ dependence of the bond variables.
The labels of the bonds are shown also.  The
length difference between the longest and shortest bonds of $\soc$
is 0.05\AA\ in the present parameters.  When $N \sim 500$, the value
becomes about one-tenth of it.  This would be smaller than the maximum value
(about 0.01\AA) of the dimerization strength observable in experiments.
In the narrowest tubules actually produced, the maximum
length is about 1000 times of that of the tubule diameter.
Such the tubules can be regarded as infinitely long.  The extrapolated
value of the length difference is $2.64 \times 10^{-3}$\AA.
Even though this value would not be observable directly,
the Kekul\'{e}-type fluctuations might survive thermal
fluctuations if we look at the interbond correlation functions.

We have also calculated about the tubule (8,2).  The energy gap $\eg$ and
the magnitudes of the bond variables are not so much different from
those of the tubule (5,5), quantitatively.  The number of carbons
arranged around the axis is ten for both tubules.  This indicates that
the electronic and lattice properties are mainly determined by the
diameter of tubules.  They do not sensitively depend on whether
the tubule is helical or not.

We have looked at the variation of electronic and lattice structures
from $\soc$ and $\rug$ to an infinitely long tubule (5,5) in [4].
Ten more carbons have been inserted successively.
The linear $1/N$-dependence is similarly found in the data of $\eg$ and
the bond variables.  We have concluded the Kekul\'{e} pattern in
the lattice structure, too.

\vspace{0.5cm}
\noindent
{CHAIN-LIKE DISTORTION}

We discuss about the tubules, (6,4) and (7,3).  In these tubules, the
Kekul\'{e} pattern is automatically excluded owing to the boundary
condition.  We have obtained the solutions where
\underline{the {\sl trans}-polyacetylene
chains are arranged along almost the tubular axis}.  We call this pattern
``chain-like distortion''.

Figure 4 shows $\eg$ of the tubule (6,4) as a
function of $1/N$.  The energy gap
almost saturates at $N \sim 600$.  The large gap ($\sim$ 1eV) remains
when $N \rightarrow \infty$.  This is due to the fact that there
is a wide gap even in the system with $\lambda = 0$.  The system
is a \underline{semiconductor} whether there is a dimerization
pattern or not.

In Fig. 5, we show the bond variables against $1/N$.   The label of
each bond is shown, too.  It is apparent that there is a dimerization
pattern in the infinitely long tubule.  However, the strength of the
dimerization is very small: the length difference between the shortest
and the longest bonds is about 0.003\AA.  This value is of the
magnitude similar to that found in Fig. 3.  It may be mainly determined by
the number of carbons perpendicular to the axis.

For the tubule (7,3), we have obtained the same kind of solutions.
The bond alternation pattern is chain-like.
The energy gap and the strength of the dimerization do not
change so much from those in Figs. 4 and 5.

\vspace{0.5cm}
\noindent
{DISCUSSION}

We have investigated the tubules where
ten carbons are arranged in the direction
perpendicular to the tubular axis.  We have obtained the similar strength
of dimerizations for the Kekul\'{e} structure and the chain-like
distortion.  The strength is about one order smaller than the experimentally
accessible magnitude.  Therefore, it would be difficult to observe
directly the bond alternation patterns in the very long tubules.
However, the fluctuations of the phonons from the classical values
might show some correlations which reflect the Kekul\'{e} or
chain-like patterns.

\vspace{0.5cm}
\noindent
{REFERENCES}

\noindent
1 S. Iijima, \underline{Nature, 354} (1991) 56; T. W. Ebbesen and P. M.
Ajayan, \underline{Nature, 358} (1992) 220.\\
2 N. Hamada, S. Sawada, and A. Oshiyama, \underline{Phys. Rev. Lett., 68}
(1992) 1579; R. Saito, M. Fujita, G. Dresselhaus, and M. Dresselhaus,
\underline{Phys. Rev. B,} (in press); K. Tanaka, M. Okada, K. Okahara,
and T. Yamabe, \underline{Chem. Phys. Lett., 193} (1992) 101.\\
3 W. P. Su, J. R. Schrieffer, and A. J. Heeger,
\underline{Phys. Rev. B, 22} (1980) 2099.\\
4 K. Harigaya, \underline{Phys. Rev. B, 45} (1992) 12071.\\

\noindent
Fig. 1. Possible way of making helical and nonhelical tubules.
The open and closed circles indicate the metallic and semiconductoring
behaviors of the undimerized system, respectively.  The Kekul\'{e}
structure is superposed on the honeycomb lattice pattern.

{}~

\noindent
Fig. 2.  $1/N$ dependence of the energy gap $\eg$ of the tubule (5,5).

{}~

\noindent
Fig. 3.  $1/N$ dependence of the bond variables of the tubule (5,5).
The labels of bonds are shown also.

{}~

\noindent
Fig. 4.  $1/N$ dependence of the energy gap $\eg$ of the tubule (6,4).

{}~

\noindent
Fig. 5.  $1/N$ dependence of the bond variables of the tubule (6,4).
The labels of bonds are shown also.

\noindent
Note: In order to obtain figures, please contact with the following
address: harigaya@etl.go.jp.  The figures will be sent by the
conventional mail.
\end{document}